\documentclass[letterpaper,english,reprint, aps]{revtex4-1}
\usepackage[T1]{fontenc}
\usepackage[latin9]{inputenc}
\usepackage{amsmath}
\usepackage{amssymb}
\usepackage{graphicx}
\usepackage{esint}



\usepackage{babel}
\begin{document}

\preprint{APS/123-QED}

\title{The mutual energy current interpretation for quantum mechanics}

\author{Shuang-ren Zhao}

\affiliation{Imrecons Inc}

\author{Kevin Yang}

\affiliation{Imrecons Inc}

\author{Kang Yang}

\affiliation{Imrecons Inc}

\author{Xingang Yang}

\affiliation{Imrecons Inc}

\author{Xingtie Yang}

\affiliation{Northwestern Polytechnical university}

\date{\today}
\begin{abstract}
Quantum physics has the probability interpretation. Traditionally
we have believed the particle for example electron looks like the
light wave. From the knowledge of light, we know that wave is always
spread out, and hence the electron wave should also spread out. That
means the electron wave beam should like the light wave beam become
diverged from the source. When the electron is received by an atom
we thought the wave collapse. The place to collapse is depends on
the probability calculated from the square of absolute value of the
wave function. The recent new discovery tell us that the light is
not just wave, it is a combination of waves, retarded potential and
advanced potential. These two potentials together produce the mutual
energy current or referred as M-current. Hence light is not a wave
and not particles, it is M-current. Light energy current is often
described as a surface integral of Poynting vector. This energy current
can be referred as P-current. The new discovery found that P-current
doesn't carry any energy for light. The contribution of P-current
to energy transfer can be omitted. The light energy is transferred
only by M-current. The beam of M-current doesn't like the beam of
P-current which is diverged from the source, instead, the M-current
beam first diverges from the source (a electron in an atom) and then
converged to the sink (an electron in another atom). Since the M-current
at the place to be received is localized at one electron, the concept
of wave function collapse is needless. The probability results of
light is because that we have use P-current to roughly calculate the
M-current. We thought if Schrödinger knew today's light theory, he
would for sure also build his wave theory for quantum mechanics similar
to the new light theory with M-current. Hence we claim that the M-current
theory is not only suitable to the light but also can be applied to
the quantum physics. This means all particles are M-current. The M-current
is composed of not only the retarded potential, but also the advanced
potential. M-current is a inner product of the retarded potential
and the advanced potential. Since there is two waves, the M-current
theory can offer a very nature way to the phenomena of spin. The traditional
wave energy current which is calculated with only retarded potential,
we obtained the P-current. If we still calculate P-current for particles,
we have to use the concept of probability, however P-current is only
an inaccurate calculation of the energy current. More accurate method
should be the M-current method. In this new method, the concept of
the wave function collapse is not necessary. We still can use the
probability, however the reason of the probability becomes very clear.
\begin{description}
\item [{Usage}]~{\small \par}
\item [{PACS~numbers}] 52.25.Os, 41.20.Jb, 14.70.Bh, 25.20.Dc, 03.65.-w,
12.20.-m, 42.50.-p{\small \par}
\item [{Structure}]~{\small \par}
\end{description}
\end{abstract}



\maketitle

\section{Introduction}

In quantum physics, particle for example electron must satisfy the
wave function. The wave is a retarded wave. This wave is a diverged
wave which spread out from the source to a bigger region. But when
electron is absorbed by an atom, we say that the wave function collapses.
Hence it seems that the square absolute value of the wave function
can offer the probability of the position of wave function collapse.
This become the probability interpretation of quantum physics, it
is also referred as the Copenhagen interpretation\citep{IEEEexample:HermannWimmel}.
This interpretation cannot be accepted by many physicists. Einstein
said ``God does not play at dice.'' To which Bohr could only answer:
``But still, it cannot be for us to tell God, how he is to run the
world''. Schrödinger disagreed also the probability interpretation
of the quantum physics. There are other interpretation for example
many-worlds interpretation\citep{IEEEexample:Everett} and consistent
histories\citep{IEEEexample:Griffiths} now they become also very
popular. The transactional interpretation\citep{IEEEexample:Wheeler_1,IEEEexample:Wheeler_2,IEEEexample:JohnCramer1,IEEEexample:JohnCramer2}
is interested with more and more physicists. However until now seems
still no any other interpretations can replace the probability interpretation.

In light situation, traditionally we know there is light wave which
is retarded potential, from this wave we can calculate the Poynting
vector, a surface integral of Poynting vector can be seen as energy
current, it is referred as P-current. P-current looks like a diverged
beam started from the source go to the surrounding environment.

Recently the authors' study\citep{IEEEexample:shrzhao1,IEEEexample:shrzhao2,IEEEexample:shrzhao3,IEEEexample:ConceptMutualEnergy,IEEEexample:shuangrenzhaoarxiv_2}
shown that light is not just wave, the retarded potential. Light is
mutual energy current. Mutual energy current is referred as M-current.
M-current is composed of the retarded potential and advanced potential.
M-current is a inner product of the two potentials. Interesting thing
is the two different potentials can be synchronized to transfer the
energy. The beam of M-current in the beginning diverges from a electron
in a atom which can be seen as the source, and then it converged to
another electron which can been seen as sink. According this new theory
of light, since M-current itself can be localized, the concept of
the wave function collapse is not necessary. Since there are two potentials,
the phenomena of spin can also be easily explained as two potentials
with 90 degree phase difference.

The authors believe that all result about M-current in light is also
suitable to other particles for example electrons. The authors believe
that the electron is actually the M-current, which is composed of
two waves, retarded wave and advanced wave. In the orbit of atom,
the two waves can be synchronized completely, this is similar to the
light transfer in a wave guide. In this situation, the behavior of
two waves is same to the behavior obtained with the retarded wave
alone. Similar to light, the energy current is calculated with only
retarded wave alone (or advanced wave alone) is referred as P-current.
In free space, the behavior of M-current is different with P-current.
P-current is a diverged beam like the beam of flashlight. However
M-current started from an electron in an atom and end at an electron
at another atom. The beam M-current first diverges from a point and
then converges to a point. M-current can also interpret the phenomena
of spin in a nature way. The authors' above theory can be referred
as mutual energy current interpretation.

Compare to the transactional interpretation\citep{IEEEexample:Wheeler_1,IEEEexample:Wheeler_2,IEEEexample:JohnCramer1,IEEEexample:JohnCramer2},
the mutual energy current interpretation also support the existent
of the advanced wave. The transactional interpretation claim it is
only a interpretation, the advanced wave can not be tested in the
laboratory and also can not be falsified. However the mutual energy
current interpretation is not just a interpretation, it also changed
the quantum theory, the inner product of advanced wave and retarded
wave produced the M-current. Because of this inner product, the beam
shape of the M-current become the intersection of sets of the beam
shapes of the advanced wave and the retarded wave. This beam shape
become first diverged and then converged. In other hand, the authors
have proven that the mutual energy theory is agreed with energy conservation
law, but Lorentz reciprocity theory doesn't in case of the loss media
in the electromagnetic field theory\citep{IEEEexample:shuangRenZhaoNewTestmony}.
The mutual energy theorem and Lorentz reciprocity theory can solve
same problem for example to find the direction pattern of receiver
antenna. The mutual energy theorem needs advanced potential but the
Lorentz reciprocity theorem needs only retarded potential. 

\section{Review M-current in electromagnetic fields}

\subsection{Introduction of the M-current in light and electromagnetic field}

Maxwell equations have two solutions: the retarded potential and the
advanced potential. Many physicists accept the advanced potential,
for example the theory of Lawrence M. Stephenson\citep{IEEEexample:LawrenceMStephenson}.
The absorb theory of D. T. Pegg, Wheeler and Feynman \citep{IEEEexample:Wheeler_1,IEEEexample:Wheeler_2,IEEEexample:Pegg,IEEEexample:JohnCramer1,IEEEexample:JohnCramer2}
also based on the advanced potential. Einstein has joined a talk of
Feynman about his absorb theory that need a complete absorber in the
future. After the talk, Pauli object the theory and ask Einstein to
agree to him. Einstein did not agree that, he support the absorb theory
and said, without the absorb theory, it would be very difficult to
make a corresponding theory for gravitational interaction. Actually
in 1909 Einstein has debated with Ritz about advanced potential. Ritz
has a emission theory based on only retarded potential, hence he was
against the advanced potential. Einstein applaud the concept of advanced
potential. Although there are so many important physicists applaud
the concept of the advanced potential, but there is no any selfconsistent
theory can apply advanced potential in our electromagnetic field calculation.
How advanced potential worked in antenna system, power station system
and electronic circuits? No one offers a convinced answer. Hence,
antenna and microwave engineers, electronic engineers and most physicists
still reject the concept of the advanced potential. 

It seems without advanced potential the engineer still can solve all
engineering problems by using the reciprocity theorem\citep{IEEEexample:Lorentz-1,IEEEexample:Carson-1,IEEEexample:Carson-2,IEEEexample:STUART-BALLANTINE,IEEEexample:Rumsey}\citep{IEEEexample:JinAuKong,IEEEexample:J_A_Kong2}. 

The author introduced the concept of mutual energy and the mutual
energy theorem\citep{IEEEexample:shrzhao1,IEEEexample:shrzhao2,IEEEexample:shrzhao3,IEEEexample:ConceptMutualEnergy}.
There were very closed earlier publications related to the mutual
energy theorem\citep{IEEEexample:Welch}\citep{IEEEexample:Rumsey_VH},
but they did not realize it is the energy and still thought it as
some kind of reciprocity. The author of ref.\citep{IEEEexample:shrzhao1,IEEEexample:shrzhao2,IEEEexample:shrzhao3}
thought it is energy and referred it as ``mutual energy'' but didn't
continue to work on it for a long time. The reason is that there is
another energy current which is based on an surface integral of Poynting
vector, it is referred as P-current. The mutual energy current (it
is referred as M-current) is only part of P-current. Another reason
is theory of the mutual energy is based on the mutual energy theorem
which can be used to the calculation of the antenna system with at
least one receive antenna. However there is the reciprocity theorem\citep{IEEEexample:Lorentz-1,IEEEexample:Carson-1,IEEEexample:Carson-2,IEEEexample:STUART-BALLANTINE,IEEEexample:Rumsey}\citep{IEEEexample:JinAuKong,IEEEexample:J_A_Kong2}.
Which is also can be used to do the same thing and have been widely
applied. In the mutual energy theorem advanced potential is involved,
that is weird. In the reciprocity theorem, we do not need advanced
potential. Only recently the authors have found that the contribution
of P-current to light energy transfer can be omitted \citep{IEEEexample:shuangrenzhaoarxiv_2,IEEEexample:ShuangrenZhaoMcurrent}.
Hence M-current becomes the only one that can carry energy. It is
the only way to transfer energy between two remote objects. By the
way, to transfer energy in wave guides or coaxial-cables, energy transferred
by M-current and P-current is half to half. In antenna system the
M-current is dominant, but P-current still have some contribution.
Recently the authors have also shown that the in the loss media, the
reciprocity theorem cannot offer any results which guarantee the energy
conservation, but mutual energy theorem can do\citep{IEEEexample:shuangrenzhaoarxiv_2}.
After these two breakthroughs, it is clear to the authors that light
should be explained as M-current and light is not only the retarded
potential, it is a combination of a retarded potential and an advanced
potential. It is a inner product of the retarded potential and an
advanced potential. The authors found that the duality of the light,
photon and wave both can be explained with M-current. M-current diverges
in the beginning and then it converges to a point and hence M-current
is local which eliminate the concept of the wave function collapse.
M-current can also interpret the phenomena of spin in a nature way.

\subsection{The mutual energy theorem}

\subsubsection{Mutual energy theorem in loss media}

The theory for the mutual energy theorem includes the following components.
Before the full formula of the mutual energy theorem, there are two
early version of it, which can be seen in\citep{IEEEexample:Welch}\citep{IEEEexample:Rumsey_VH}.
The formula of the mutual energy theorem can be found \citep{IEEEexample:shrzhao1,IEEEexample:shrzhao2,IEEEexample:shrzhao3}.
The formula \citep{IEEEexample:shrzhao1,IEEEexample:shrzhao2,IEEEexample:shrzhao3}
is in Fourier domain. The corresponding time domain mutual energy
theorem can be found in ref.\citep{IEEEexample:Adrianus2}. 

In Fourier domain the modified mutual energy theorem formula can be
written as following\citep{IEEEexample:ConceptMutualEnergy},
\begin{equation}
(\xi_{1},\xi_{2})_{\varGamma}+(\rho_{1},\xi_{2})_{V}+(\xi_{1},\rho_{2})_{V}=0\label{eq:2000-130-1}
\end{equation}
$V$ is the volume contains the current $\rho_{1}=[J_{1},K_{1}]$
and $\rho_{2}=[J_{2},K_{2}]$. $J_{1},J_{2}$ are electric current,
$K_{1},K_{2}$ are magnetic current. $\varGamma$ is the boundary
of the volume $V$. $\varGamma$ can be chosen as infinite big sphere.
The media have to meet the condition, $\xi_{1}=[E_{1},H_{1}]$, $\xi_{2}=[E_{2},H_{2}]$,
$E_{1},E_{2}$ are electric fields, $H_{1},$ $H_{2}$ are magnetic
fields. Here the media $\epsilon_{1},\mu_{1}$ and $\epsilon_{2},\mu_{2}$
satisfy,
\begin{equation}
\epsilon_{1}^{\dagger}(\omega)=\epsilon_{2}(\omega),\ \ \ \ \mu_{1}^{\dagger}(\omega)=\mu_{2}(\omega)\label{eq:2000-150-1}
\end{equation}
Where $\epsilon$ and $\mu$ are permittivity, permeability, ``$\epsilon^{\dagger}=\epsilon^{*T}$.
The superscript$*$ expresses the complex conjugate operator, $T$
is matrix transpose and 
\begin{equation}
(\xi_{1},\xi_{2})_{\varGamma}\equiv\varoiintop_{\varGamma}\,(E_{1}\times H_{2}^{*}+E_{2}^{*}\times H_{1}^{*})\cdot\hat{n}dS\label{eq:2000-150-2}
\end{equation}
\begin{equation}
(\rho_{1},\xi_{2})_{V}\equiv\iiintop_{V}(E_{2}^{*}\cdot J_{1}+H_{2}^{*}\cdot K_{1})\,dV\label{eq:2000-150-3}
\end{equation}
\begin{equation}
(\xi_{1},\rho_{2})_{V}\equiv\iiintop_{V}(E_{1}\cdot J_{2}^{*}+H_{1}\cdot K_{2}^{*})\,dV\label{eq:2000-150-4}
\end{equation}
The symbol $\equiv$ means ``is defined as''. It is possible that
the modified mutual energy theorem is not a physical theorem since
the media of the two fields $\zeta_{1}$ and $\zeta_{2}$ can be different
or at different spaces. If we assume they are the same, i.e.,$\epsilon_{1}=\epsilon_{2}$
and $\mu_{1}=\mu_{2}$. There is
\begin{equation}
\epsilon^{\dagger}(\omega)=\epsilon(\omega),\ \ \ \ \mu^{\dagger}(\omega)=\mu(\omega)\label{eq:2000-154}
\end{equation}
That is lossless condition. Hence in lossless media the mutual energy
theorem (note here, there is no ``modified'') is established. Eq.(\ref{eq:2000-130-1})
can be written as
\begin{equation}
(\xi_{1},\xi_{2})_{\varGamma}=-(\rho_{1},\xi_{2})_{V}-(\xi_{1},\rho_{2})_{V}\label{eq:2000-130-2}
\end{equation}
The left side of the formula is the energy current send to outside
of the big sphere. The right side are the emission energies from the
sources(or sinks) $\rho_{1}$ and $\rho_{2}$. $(\rho_{1},\xi_{2})_{V}$
can be seen as received energy. $-(\rho_{1},\xi_{2})_{V}$ can be
seen as the emission energy. 

If the media has energy loss, It can be proven that the emission energy
should equal to the summation of the outgo energy and the energy loss,
\begin{equation}
(\xi_{1},\xi_{2})_{\varGamma}+Q_{loss}=-(\rho_{1},\xi_{2})_{V}-(\xi_{1},\rho_{2})_{V}\label{eq:2000-130-3}
\end{equation}
Here $Q_{loss}$ is the energy loss in the media. Hence the above
formula is referred as the mutual energy theorem with energy loss. 

\subsubsection{M-current is inner product}

The surface integral in the mutual energy formula $(\xi_{1},\xi_{2})_{\varGamma}$
is a inner product \citep{IEEEexample:shrzhao1}. Here we assume that
that the two fields $\xi_{1},\xi_{2}$ are two retarded potentials,
the surface integral satisfies following 3 inner product laws,

I. conjugate symmetry,
\begin{equation}
(\xi_{1},\xi_{2})_{\varGamma}=(\xi_{2},\xi_{1})_{\varGamma}^{*}\label{eq:6010}
\end{equation}

II linear,
\begin{equation}
(\xi_{1}+\xi_{2},\xi_{3})_{\varGamma}=(\xi_{1},\xi_{3})_{\varGamma}+(\xi_{2},\xi_{3})_{\varGamma}\label{eq:6020}
\end{equation}
\begin{equation}
(\alpha\xi_{1},\xi_{2})_{\varGamma}=\alpha(\xi_{1},\xi_{2})_{\varGamma}\label{eq:6030}
\end{equation}

III Positive-definiteness
\begin{equation}
(\xi,\xi)_{\varGamma}\geq0\label{eq:6040}
\end{equation}
\begin{equation}
(\xi,\xi)_{\varGamma}=0\ \ \ \ \ \ \ \iff\ x=0\label{eq:6050}
\end{equation}
Here $\iff$ means if and only if. If the two fields are all advanced
potentials, the most above results can also be obtained except,

\begin{equation}
(\xi,\xi)_{\varGamma}\leq0\label{eq:6040-1}
\end{equation}

In the mixture situation, If we $\xi=\xi_{1}+\xi_{2}$, and $\xi_{1}$
one is retarded potential and $\xi_{2}$ is advanced potential, the
above most inner product laws are still satisfied, except the formula
Eq.(\ref{eq:6040},\ref{eq:6050}) does not satisfy. Even so, this
still means the above surface integral is a good inner product. The
above inner product formulas guarantee that we can use the inner product
expression $(\xi_{1},\xi_{2})_{\varGamma}$. Mutual energy current
is referred as M-current. In optics vector field become scale field.
M-current is a inner product, that means, we can image $\xi_{1}$
and $\xi_{2}$ as two scale value, and hence $(\xi_{1},\xi_{2})_{\varGamma}\rightarrow\xi_{1}\xi_{2}^{*}$
in optics and quantum mechnics. 

Assume $\zeta_{1}$ is retarded potential and $\zeta_{2}$ is advanced
potential we can prove that in the free space (where the media is
$\epsilon_{0}$ and $\mu_{0}$), it can be proven that the surface
integral of the mutual energy theorem will vanish at infinite big
sphere $\varGamma$
\begin{equation}
\lim_{r\rightarrow\infty}(\xi_{1},\xi_{2})_{\varGamma}=0\label{eq:2000-160}
\end{equation}
where $r$ is the radio of the surface $\varGamma$.

\subsection{The system with transmit antenna and receive antenna\label{sec:The-system-withVII}}

We have proven that in the mutual energy theorem, if $\xi_{1}$ and
$\xi_{2}$ one is retarded potential and another is advanced potential,
the surface integral vanishes, i.e.,$(\xi_{1},\xi_{2})_{\varGamma}=0$.
Here $\varGamma$ is infinite big sphere. $\hat{n}$ is the outward
unit vector of surface $\varGamma$. This means no mutual energy can
go out of the big sphere $\varGamma$. The mutual energy can only
go from $\rho_{1}$ to $\rho_{2}$. Assume, the media ($\epsilon,$$\mu$)
are loss less, in that situation there is $Q_{loss}=0$, we have
\begin{equation}
(\xi_{1},\rho_{2})_{V_{1}}=-(\rho_{1},\xi_{2})_{V_{1}}\label{eq:4040}
\end{equation}
Here $-(\rho_{1},\xi_{2})_{V_{1}}$ is the emitted energy of $\rho_{1}$,
$(\xi_{1},\rho_{2})_{V_{2}}$ is the received energy of $\rho_{2}$.
If we choose a surface $\varGamma_{1}$ in which, there is only $\rho_{1}$
inside and $\rho_{2}$ is at outside, the mutual energy theorem can
be written as
\begin{equation}
(\xi_{1},\xi_{2})_{\varGamma_{1}}=-(\rho_{1},\xi_{2})_{V_{1}}\label{eq:4041}
\end{equation}
$(\xi_{1},\xi_{2})_{\varGamma_{1}}$ is the mutual energy current
or M-current between the source $\rho_{1}$ and the sink $\rho_{2}$.
$\varGamma_{1}$ can be any surface between the two current $\rho_{1}$
and $\rho_{2}$. M-current is equal to the emitted energy $-(\rho_{1},\xi_{2})_{V_{1}}$
and the received energy $(\xi_{1},\rho_{2})_{V_{2}}$.

\subsection{The beam shape of the M-current.}

\begin{figure}
\includegraphics[scale=0.4]{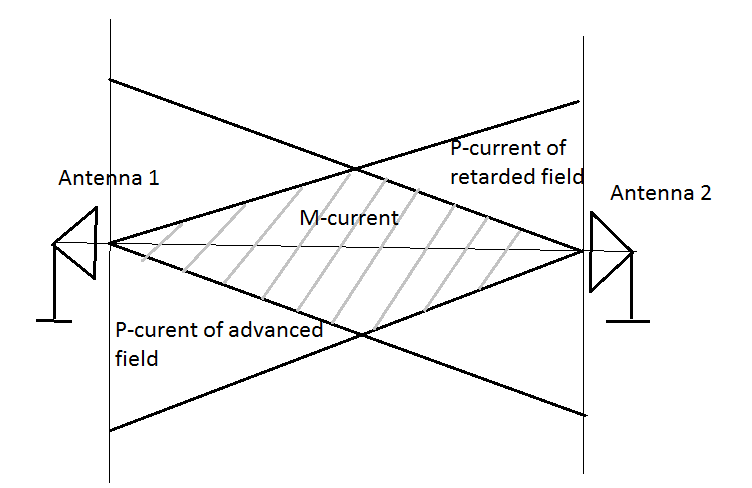} 

\caption{Antenna system with a transmitter and a receiver, the transmitter
send the retarded potential, this potential will cause the current
change in the receiver and the receiver will cause an advanced potential.
The shape of beam of P-current is diverged. The shape of beam of M-current
first diverges and then converges which is shown in a dashed area.\label{fig:Antenna-system-with}}
\end{figure}

Assume $\xi_{1}$ is retarded potential, $\xi_{2}$ is advanced potential,
\[
(\xi_{1},\xi_{1})_{\varGamma_{1}}=\varoiintop_{\varGamma_{1}}(E_{1}\times H_{1}^{*}+E_{1}^{*}\times H_{1})\cdot\hat{n}d\varGamma
\]
\begin{equation}
=2\,\varoiintop_{\varGamma_{1}}\Re\{E_{1}\times H_{1}^{*}\}\cdot\hat{n}d\varGamma=2\varoiintop_{\varGamma_{1}}\Re\{S_{1}\}\cdot\hat{n}d\varGamma\label{eq:4043}
\end{equation}
where $\Re\{\bullet\}$ means take the real values. $\varGamma_{1}$
is any surface between $\rho_{1}$ and $\rho_{2}$. $S_{1}=E_{1}\times H_{1}^{*}$
is Poynting vector, hence $(\xi_{1},\xi_{1})_{\varGamma_{1}}$ is
P-current. The beam shape of P-current is similar to the that of retarded
potential $\xi_{1}$, which is diverged from the source $\rho_{1}.$
Similarly that $(\xi_{2},\xi_{2})_{\varGamma}$ is also P-current
the beam shape of it is diverged from the sink $\rho_{2}$. 

\[
(\xi_{2},\xi_{2})_{\varGamma_{2}}=\varoiintop_{\varGamma_{2}}(E_{2}\times H_{2}^{*}+E_{2}^{*}\times H_{2})\cdot\hat{n}d\varGamma
\]
\begin{equation}
=2\,\varoiintop_{\varGamma_{2}}\Re\{E_{2}\times H_{2}^{*}\}\cdot\hat{n}d\varGamma=2\varoiintop_{\varGamma_{2}}\Re\{S_{2}\}\cdot\hat{n}d\varGamma\label{eq:4043-1}
\end{equation}
$\varGamma_{2}$ is any surface between $\rho_{2}$ and $\rho_{1}$.
The beam shape of the $(\xi_{1},\xi_{2})_{\varGamma}$ mutual energy
current can be obtained largely according to 
\begin{equation}
beam\{(\xi_{1},\xi_{2})_{\varGamma}\}=beam\{(\xi_{1},\xi_{1})_{\varGamma_{1}}\}\cap beam\{(\xi_{2},\xi_{2})_{\varGamma_{2}}\}\label{eq:4044}
\end{equation}
where $beam\{\bullet\}$ shows the beam shape of the energy current.
The symbol $\cap$ means the intersection sets. Hence the beam shape
of M-current in the beginning diverges from the source $\rho_{1}$
and then converges to the sink $\rho_{2}$. The beam shape of M-current
can be seen in Figure \ref{fig:Antenna-system-with}.

\subsection{Retarded potential and advanced potential can be synchronized}

We often think the retarded potential and advanced potential are totally
different things, but actually they looks similar even can be synchronized.
For example in the wave guide, assume in the two ends there is a transmitter
and a receiver, the wave between the two ends can be seen as retarded
potential to the transmitter and also can be seen as advanced potential
to the receiver. In the transactional interpretation\citep{IEEEexample:JohnCramer1,IEEEexample:JohnCramer2},
they think the advanced potential can not be tested in laboratory
and it is impossible to be falsified. We do not agree with them. We
believe advanced potential can be tested we will discuss this a future
article. Here we can see the advanced potential sent by the receiver
indeed is synchronized with the retarded potential sent by the transmitter.

We assume in the wave guide, the transmitter contributed half retarded
potential the receiver contributed another half. But even the two
half are retarded potential and advanced potential they are completely
synchronized, hence, there is,
\begin{equation}
\xi_{1}(t,x)=\xi_{2}(t,x)=\frac{1}{2}\xi(t,x)\label{eq:4045}
\end{equation}

\[
\xi(t,x)=\xi_{1}(t,x)+\xi_{2}(t,x)
\]
where $t,x$ are time and position in the wave guide. Traditionally
we assume in the wave guide there is only retarded potential, even
this is wrong, but all results are still correct.

In other situation, for example two antennas one is the transmitter
and another is the receiver. The beam shape of the retarded potential
and advanced potential are different, one is diverged from the source,
the other is diverged from the sink, but the value inside the M-current
beam is still very good synchronized with time and position. This
is also the reason why the M-current can transfer energy.

\section{Review M-current for light}

\subsection{Photo}

If we assume the energy emitted from source $\rho_{1}$ is discrete
and the sink $\rho_{2}$ receive energy also discrete. All electrons
in an atom in the environment can be seen as sinks which can absorb
the energy. Assume the every electron can randomly jump from low energy
to the high energy and also can randomly jump from low energy to high
energy. The probability of the jump is very low. Assume in the environment
there is a absorber electron from lower energy jump to high energy,
hence sends advanced potential to the transmitter. In the same time
in the emitter there is a electron jump from high energy down to lower
energy. If the time just match each other, the retarded potential
of the emitter and the advanced potential of the abosorber build a
M-current, which send the energy from the emitter to the abosorber.
In this case of the mutual energy current, the electron in the absorber
stayed at higher energy. The electron in the emitter kept at the lower
energy. The electron in the abosorber keep in the higher energy. If
in the above the two time windows does not match, the mutual energy
current does not build. If there has not built the mutual energy current,
the electron at absorber will return to its original energy lever
and the electron at the emitter will also return to its original energy.

\begin{figure}
\includegraphics[scale=0.4]{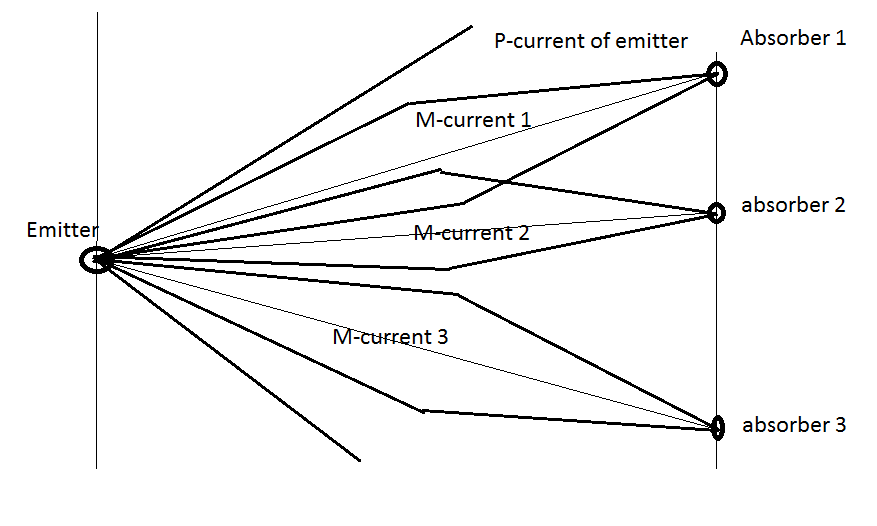} 

\caption{The M-current of 3 Photons and P-current of the emitter.\label{fig:Antenna-system-with-light}}
\end{figure}

In another situation, we have assumed that for the electron in absorber
there is only a very short time windows which can absorb the retarded
potential. There will be randomly some electron react to the retarded
potential emitted from $\rho_{1}$. The M-current can be built between
the source and the sinks. Since this M-current is localized in the
place it is absorbed, it looks a particle when it is received and
emitted. In place between the source and a sink, the beam of M-current
is much wider, and looks very like waves. This is why when light go
through the double slit will looks like wave, but in the absorb place
it becomes many points and looks like particles, which can be seen
as photons. Figure \ref{fig:Antenna-system-with-light} shows the
tree absorbers reacted with the emitter, hence there are 3 M-currents
transferred from the emitter to the absorbers, these can be seen as
3 photons. The figure shows that the shape of M-currents are focused
at absorbers, there is no so called wave collapse in the absorbers. 

\subsection{Double slits}

\begin{figure}
\includegraphics[scale=0.3]{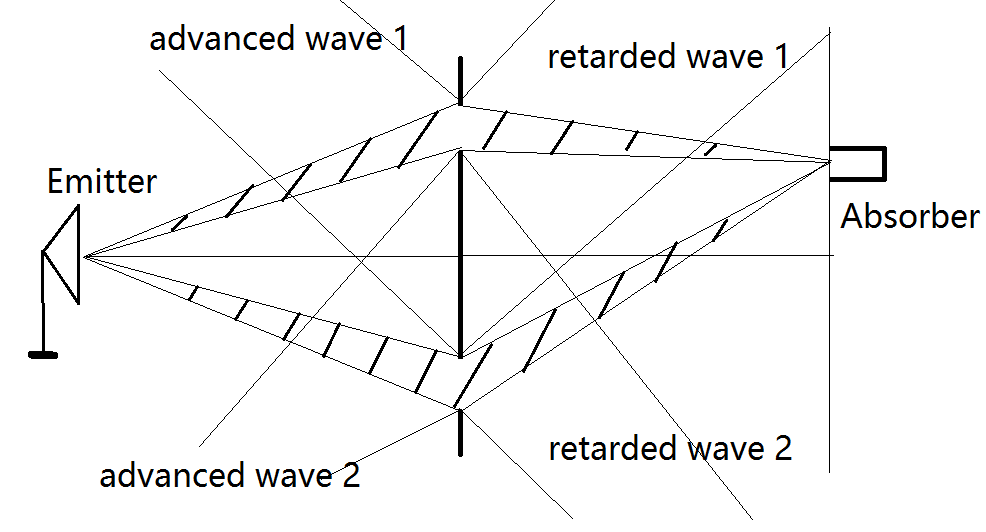} 

\caption{The M-current of 3 Photons and P-current of the emitter.\label{fig:double-slit-3}}
\end{figure}

For light we must explain the experiment of double slits. The mutual
energy current method offers a accurate method to calculate the energy
go through the slits. The retarded potential and advanced potential
can all be calculated through the Huygens principle. There is the
version modified version of Huygens principle which is base on the
mutual energy theorem\citep{IEEEexample:shrzhao2}. The second step
is calculate the mutual energy which is the inner product of the advanced
potential and retarded potential. The mutual energy theorem should
help us exactly calculate the intensity of the electromagnetic fields.
However we can easy to find out the beam shape of the mutual energy
current which can be seen in Figure \ref{fig:double-slit-3}.

We can see the light energy which is M-current goes through double
slits, in the end the light line combined together to the place of
the absorber. It is clear there is no need of wave function collapse.
Compare to the transactional interpretation\citep{IEEEexample:Wheeler_1,IEEEexample:Wheeler_2,IEEEexample:JohnCramer1,IEEEexample:JohnCramer2},
since the both retarded potential and advanced potential are still
separated, hence the wave function double collapses are needed for
both retarded and advanced potential in the absorber and emitter.
Transactional interpretation can not offer energy current calculation.
M-current method not only offer a interpretation but also offer a
very accurate method to calculate the energy current in any time and
surface. 

It is clear that when light go through the double slits, it looks
like light will produce the interference pattern. In the place light
is absorbed, like looks light particle, because all M-current energy
has focused to a point. 

\subsection{Delay choice}

The above M-current model is also very easy to explain the delay choice
experiment\citep{IEEEexample:R_Marlow,IEEEexample:WheelerDelayedChoice}.
Because the advanced potential is send out from absorber, even the
photon (light or M-current) has go through the double slits, if the
absorber configuration is changed it is still can change the beam
shape of the M-current. For example if at the time we know the photon
(light or M-current) has go through the double slits, if a calibrator
is quickly added to the front of the absorber which direct to one
of slit, the beam shape of M-current is changed, only one side of
M-current up or down left, and hence the light is go only from one
slit. The photon looks as particle.

From the view of M-current, the concept of photon, is not complete
correct. In the space light can be as M-current go through the double
slit. It is focused to a point only when it is emitted or absorbed.
However since this energy is still discrete, we can still call it
as photon. If we know the photon is just M-current, it is very easy
to understand the behave when it go through the double slits.

Delayed choice experiment tell us the observer can change the results
of experiment. This actually is very easy to understand. In the early
time, the people think the principle of eye is like today's ultrasound
machine or radar system which sends wave to the object and receiver
the return wave. In the M-current interpretation, the eye actually
just looks very like the the radar system or ultrasound system, the
only difference is the eye sends advanced potential instead of retarded
potenital. 

\subsection{Spin}

The phenomena of spin can also be explained easily. When we thought
the light is only composed of only one potential, when we found it
has the property of circle polarization, it becomes very confuse.
How can only one field produce a circle polarization? The circle polarization
needs two wave superposed together with 90 degree phase difference.
We can only say that there is a spin which is the light's intrinsic
property.

Now according to the above theory of M-current, it just have two potentials
and they are nearly synchronized. If we assume that there is 90 degree
difference between the two potentials. The light will become circle
polarized. When the receiver receives the retarded potential, its
current in the sink changes, that causes it to send an advanced potential.
In the light situation, the advanced potential has a 90 degree delay.
This 90 degree delay also can happen at the position of source. If
the sink first send an advanced potential the source reacted by send
a retarded potential a 90 degree delay. This two situations corresponding
to the two kinds of circle polarization: left and right circle polarization.
This kind of interpretation is much nature than speak about photon
spins.

\subsection{Quantum entanglement}

The photon entanglement happens when a no spin photon go through a
nonlinear media, two lower frequency photon will produced with one
is left spin and the other is right spin. We know that the angle momentum
must conserved. Hence one photon is left the other must right spin,
otherwise the angle momentum doesn't conserved. 

The quantum entanglement can also be explained with M-current. Assume
a emitter send two entanglement photons. One of the photon is received
by an absorber. This absorber will send advanced wave to the emitter.
If we receive a left/right spin photon that means the advanced potential
have a +/- 90 degree phase difference to the retarded potential. When
this event happens, the emitter will know this a time before, since
the abserver sends advanced wave. The emitter will send a right/left
photon spin photon, that means the retarded potential will have +/-
90 degree phase difference to the advanced potential.

The problem is the two photon has been send out a time, when we seen
the first one is left the second immediately (1000 time fast then
light speed) become right. 

According to M-current theory, the left/right spin of photon is not
decided by emitter, it is decided by both emitter and absorber. The
phase difference of the retarded potential of the emitter and the
advanced potential of the absorber decides the spin rotation direction.
If the absorber received a left spin photon the emitter will know
that when it send out the retarded wave. Since the second of photon
is sent also by same emitter, this emitter is a high frequency photon
in the nonlinear device with non angle momentum. When this photon
be come two low frequency photon, the angle momentum must conserved.
Hence the second low frequency photon sent by the emitter must with
different spin direction to the first low frequency photon. If we
receive the second photon we find it has different angle momentum
with the first photon. The only strange thing is the time from the
first we observer the first to the second photon, it is

\begin{equation}
\tau=\tau_{1}+\tau_{2}\label{eq:10000-500}
\end{equation}
Where $\tau$ is the time from observer the first photon, $\tau_{1}$
is the advanced wave send from first absorber to the emitter. This
time has negative value, since the wave is advanced wave. $\tau_{2}$
is the time from emitter to the second absorber, this time has positive
value. In case $|\tau_{1}|\approx|\tau_{2}|$, $\tau\approx0$. In
this case the events from first photon is absorbed to the second photon
is absorbed can have zero time. Hence this two events are not local.

\begin{equation}
\tau\approx0\label{eq:10000-510}
\end{equation}

\[
speed=\frac{L}{\tau}\gg0
\]

\section{Other particles}

In the time de Broglie and Schrödinger built their wave theory for
the particles, they do not knew the above new discoveries for light.
Hence they built their particle theory only including the retarded
wave. Now we have known for light it is M-current composed with retarded
wave and advanced wave, for us it is easy to think all particles perhaps
also have the same structure as light. The particles also transfer
energy by M-current instead of P-current. The particle's M-current
also is composed of retarded wave and advanced wave.

\subsection{Electron}

Assume electron is the M-current, which is composed of two waves,
retarded wave and advanced wave. The atom emits the electron is referred
as emitter, the atom receives the electron is referred as absorber.
The electron's mutual energy beam should similar to the mutual energy
of the light beam, it is very narrow in both emitter point and at
the absorber point and it becomes wide in the place between  the emitter
and the absorber.

Electron has fixed energy, that can be explained as the time window
of the emitter and absorbler. This time window is very narrow, only
allowing sending and receiving an amount of energy equal to a electron.
Hence there is only one or a few absorbers to randomly react the retarded
wave and send back the advanced waves. Between this particular emitter
atom and the particular absorber atom, the mutual energy is produced
which is composed with the retarded wave and the advanced wave. The
electron is just this M-current. Electron looks more like a particle
at the place it is emitted and absorbed and it look more like wave
at the place between the emitter and absorber. However electrons actually
are M-current in all the time.

\subsection{Wave function}

In the quantum physics, assume $\psi$ is the wave function, then
$|\psi|^{2}$ is explained as probability. However the authors thought
that is because the lack of the knowledge of M-current for light.
If 90 years ago Schrödinger and Dirac knew the above theory about
light they will build their quantum theory looks like light. Here
the author means the theory which explain light as M-current. The
M-current is composed of retarded wave and an advanced wave.

After we have the new understanding about light, we know that, the
situation of quantum physics should be similar to electromagnetic
wave or light wave. If we accept the advanced wave in electromagnetic
field and light, if we accept the light is just the mutual energy
current of the two waves, one is retarded wave, another is advanced
wave, we can immediately thought that in the quantum physics perhaps
there is also the advanced wave. Assume $\psi_{1},\psi_{2}$ are two
waves, we can define the M-current (mutual energy current) of quantum
physics as
\begin{equation}
Q_{12}=(\psi_{1},\psi_{2})_{\varGamma}=\varoiintop_{\varGamma}\psi_{1}\text{\ensuremath{\psi_{2}^{*}}}d\varGamma\label{eq:10000-10}
\end{equation}
Where $\varGamma$ is a surface between the emitter and absorber,
It is possible that $\psi_{1},\psi_{2}$ are vectors like in the electromagnetic
field situation, in that situation $(\psi_{1},\psi_{2})$ will define
the mutual energy current. In quantum physics $\psi_{1},\psi_{2}$
are scales (the scale is possible only a simplified version of vector
field, just like in optics we can use scale value to describe the
electromagnetic field, which actually is a vector field).

Assume in quantum physics, $\psi_{1}$ is retarded wave which send
out from a emitter atom, $\psi_{2}$ is advanced wave which is send
out from an absorber atom. When the electron is inside the orbit of
the atom, we assume there are also advanced wave and retarded wave.
This situation, is similar to electromagnetic wave in a waveguide,
the advanced wave can be completely synchronized with the retarded
wave. Traditionally we though there is only retarded wave in this
situation, but now we assume $\psi_{2}=\psi_{1}$, assume the field
of electron can be superposed, hence the field in the electron orbit
will be
\begin{equation}
\psi=\psi_{1}+\psi_{2}\label{eq:10000-20}
\end{equation}
$\psi$ is the electron field inside the orbit. Hence 
\begin{equation}
\psi_{1}=\psi_{2}=\frac{1}{2}\psi\label{eq:10000-30}
\end{equation}
\[
(\psi,\psi)=(\psi_{1}+\psi_{2})(\psi_{1}+\psi_{2})^{*}
\]
\begin{equation}
=\psi_{1}\psi_{1}^{*}+\psi_{1}\psi_{2}^{*}+\psi_{2}\psi_{1}^{*}+\psi_{2}\psi_{2}^{*}\label{eq:10000-40}
\end{equation}
$\psi_{1}\psi_{1}^{*}$ is corresponding to the retarded wave's self-energy
current. $\psi_{2}\psi_{2}^{*}$ is corresponding to the advanced
wave's self-energy current. $\psi_{1}\psi_{2}^{*}+\psi_{2}\psi_{1}^{*}$
is corresponding the M-current. $\psi_{1}\psi_{1}^{*}$ and $\psi_{2}\psi_{2}^{*}$
is P-current. P-current is not important, even we can calculate it
to obtain some values, but since it has no contribution to exchange
energy with others, hence it can be omitted. 

In this situation since $\psi_{1}=\text{\ensuremath{\psi_{2}}}$,
the calculation only with retarded wave, i.e., assume there is only
$\psi$ which is retarded wave, will obtain the same result compare
the new quantum theory with both retarded wave and advanced wave.
This is the reason why if we do not introduce the concept of advanced
wave and the M-current, quantum physics still obtains very good results
in the situation where the electron is inside an atom or inside a
potential well. This is so called static wave situation, in the static
wave the time $t$ will explicitly appear. Even in this situation
only retarded wave can also obtain good calculation results, actually
the advanced wave still exists. 

\subsection{Election in the free space}

The self-energy current $\psi_{1}\psi_{1}^{*}$, $\psi_{2}\psi_{2}^{*}$
have no contribution to the emitter atom and the absorber atom, this
is similar to the situation of light. $\psi_{1}\psi_{1}^{*}$ is a
beam diverged from the emitter, when it reach the absorber, since
the section area of the absorber atom is too small, the energy received
by absorber from the $\psi_{1}\psi_{1}^{*}$ can be omitted. $\psi_{2}\psi_{2}^{*}$
is diverged from receiver, when it reached to the emitter atom, since
the section area of the emitter is too small, this part of transferred
energy can be omitted. In this situation only the mutual energy current
is important, which is
\begin{equation}
\varoiintop_{\varGamma}(\psi_{1}\psi_{2}^{*}+\psi_{2}\psi_{1}^{*})\,d\varGamma=2\Re\{\varoiintop_{\varGamma}\psi_{1}\psi_{2}^{*}\,d\varGamma\}\label{eq:10000-50}
\end{equation}
Where $\Re\{\bullet\}$ means to take the real value. For simplification,
we will call $Q_{12}=\varoiintop_{\varGamma}\psi_{1}\psi_{2}^{*}\,d\varGamma$
as M-current. Keep in mind that the real energy between the emitter
and absorb is $2\Re{Q_{12}}$.

The mutual energy current similar to the situation of light the beam
is that the electron beam first diverged from the emitter and then
converged to the absorber. Here the emitter and the absorber are two
atoms which can send or absorb the electrons. Here since the beam
of M-current can focus to a small point, it does not need the concept
of wave function collapse. 

The wave function collapse is because we do not know there is also
the advanced wave. So we calculate $\psi_{1}\psi_{1}^{*}$ which is
a diverged beam. At the place of the wave is received, the beam of
the energy current $\psi_{1}\psi_{1}^{*}$ become very widely spread
out. When we use $\psi_{1}\psi_{1}^{*}$ as the result of quantum
physics, we have to face the wave function suddenly collapsed to a
point. After we have explained actually the electron is M-current,
the property of M-current which first diverges and then converges
can thoroughly avoid the wave function collapse. 

The probability interpretation for the wave function is because of
we only calculated from the retarded wave, $\psi_{1}\psi_{1}^{*}$,
which is inaccurate to the electron. 

The authors don't clear why this particular emitter atom connected
to another particular absorber atom. We have said it is perhaps because
just in that time, the retarded wave reached the absorber, their time
window matched together. But this is only one possibility, that is
also possible the transmitter send retarded wave includes a special
cryptograph code, which can be understand only some absorber atom.
It is also has some positive feedback between the transmitter and
the receiver that makes the connection of one pairs of atoms become
strong than others. Finally they become connected together. An electron
is sent out from the emitter atom to the absorber atom.

\subsection{Spin}

In the traditional quantum physics, there is only one wave function,
the retarded wave, when we measured some thing rotated, it is difficult
to give an explanation, hence we call it spin. However in the authors'
new quantum explanation, there are two wave functions, one $\psi_{1}$
is retarded and the other $\psi_{2}$ is advanced. The two waves are
nearly synchronized. But there is the possibility they have small
phase difference. The spin is also similar to the situation of light.
If we assume $\psi_{2}$ has 90 degree phase difference compare to
$\psi_{1}$, there is a circle polarization. Here we can assume $\psi_{1}$
and $\psi_{2}$ are transverse field one is directed $x$ axis direction
and the other is at $y$ axis direction. The wave is transfers in
$z$ axis direction. This circle polarization is the phenomena of
spin. 

In the explanation of the mutual energy current, the spin just two
waves have a 90 degree phase difference. This phase difference is
caused by the absorber atom or the emitter atom, there is a reaction
delay to their wave re-sending process. The delay happens at absorber
or at emitter will cause the phase difference as positive 90 degree
or negative 90 degree and hence there is left or right polarization.
This is phenomena is spin. 

\subsection{The Schrödinger equation considered the advanced wave}

The original Schrödinger equation which is corresponding to the retarded
wave
\begin{equation}
ih\partial\psi(x,t)=[-\frac{h^{2}}{2\mu}\nabla+V(x,t)]\psi(x,t)\label{eq:10000-60}
\end{equation}
Corresponding to the advanced wave, there is
\begin{equation}
-ih\partial\psi(x,t)=[-\frac{h^{2}}{2\mu}\nabla+V(x,t)]\psi(x,t)\label{eq:10000-70}
\end{equation}
The above is only one example to create the advanced wave, we also
can created the advanced wave using Klein-Gordon equation or Dirac
equation, or any other equation still not found, but that is beyond
the discussion of here. The point is there must have a advanced wave.

\subsection{Summary}

For a free electron, we should calculate M-current which is $\varoiintop_{\varGamma}\psi_{1}\psi_{2}^{*}d\varGamma$.
M-current is a beam first diverge and then converge, for this kind
wave beam, the concept of wave function collapse is needless. 

When the electron is inside orbit, the two wave can be synchronized
completely, and hence the retarded wave and advanced wave are equal
to each other. In this situation, the calculation of $\psi_{1}\psi_{1}^{*}$,
the P-current (we can referred it as P-current similar to the light
situation) is the same as M-current $\psi_{1}\psi_{2}^{*}$. Even
though there still exists the advanced wave, but the result is same
when we only use retarded wave to calculate, $\psi_{1}\psi_{1}^{*}=\psi_{1}\psi_{2}^{*}$.
It is same to the wave guide, in the orbit, the energy transferred
half by P-current $\psi_{1}\psi_{1}^{*}+\psi_{2}\psi_{2}^{*}$ and
half by M-current $\psi_{1}\psi_{2}^{*}+\psi_{2}\psi_{1}^{*}$. 

In the free space the contribution of P-current can be omitted completely.
Only M-current is left. Hence electron is also M-current, which is
composed of two waves retarded wave and advanced wave.

Even though the above new interpretation has not changed the calculation
of the quantum field. However because it abandons the concept of the
wave function collapse, the probability, thing become easy to understand.
Electron in the free space is nothing else, it is just M-current.
The M-current is composed of two waves one is retarded the other is
advanced. The two waves are nearly synchronized. There is 90 degree
phase difference which can be seen as the behind scene of spin. 

In the authors new interpretation the square absolute value of wave
function $|\psi|^{2}=\psi_{1}\psi_{1}^{*}$ is the P-current, which
is only a approximation to the M-current $\psi_{1}\psi_{2}^{*}$.
Since $\psi_{2}^{*}$ is difficult to obtain, there is hundreds and
thousands $\psi_{2}$ in the environment corresponding to each atom
which can receive the electron, we still only calculate $\psi_{1}\psi_{1}^{*}$.
In this situation, $\psi_{1}\psi_{1}^{*}$ offers the probability
about $\psi_{1}\psi_{2}^{*}$. 

All these can be summarized as electron is not only retarded wave,
it combines with two wave retarded and advanced. The two waves build
as M-current which is energy current instead probability. The reason
of the probability is because if we only calculated P-current of retarded
wave which can only offers the probability where the M-current can
happen. The probability is because of the inaccurate calculation.
The probability is not an intrinsic property of electron. 

We also believe the calculation of $\psi_{1}\psi_{2}^{*}$ in some
situation can replace the calculation of $\psi_{1}\psi_{1}^{*}$ and
offer more accurate results for the understanding of the electron
and all other particles. 

\section{Conclusion\label{sec:Conclusion_IX}}

Similar to the light, other particles for example electron should
also be composed with two waves, retarded wave and advanced wave.
The two waves build up the M-current. Since M-current of electron
is a beam that in the beginning diverges and in the end it converges,
hence for M-current the wave function collapse is not needed. The
reason of the probability is re-explained. 

\bibliographystyle{plainnat}
\addcontentsline{toc}{section}{\refname}\bibliography{myxampl}

\end{document}